\documentclass [prl,aps,floats,twocolumn,showpacs,groupedaddress,floatfix,amsmath]{revtex4}
\def\bc{\begin{center}}
\def\ec{\end{center}}
\def\beq{\begin{equation}}
\def\eeq{\end{equation}}
\def\bw{\begin{widetext}}
\def\ew{\end{widetext}}
\def\bea{\begin{eqnarray}}
\def\eea{\end{eqnarray}}

\def\ra{\rangle}

\usepackage{graphicx}
\usepackage{dcolumn}
\usepackage{bm}

\begin{document}
\title{Testing the topological nature of the fractional quantum Hall  
edge}
\author{Shivakumar Jolad and Jainendra K. Jain}
\affiliation{Department of Physics, Pennsylvania State University,
University Park, PA 16802}

\date{\today}

\begin{abstract}
We carry out numerical diagonalization for much larger systems than  
before by restricting the fractional quantum Hall (FQH) edge  
excitations to a basis that is exact for a short-range interaction  
and very accurate for the Coulomb interaction. This enables us to  
perform substantial tests of the predicted universality of the edge  
physics.  Our results provide compelling evidence that the behavior  
of the FQH edge is intrinsically nonuniversal, even in the absence of  
edge reconstruction, and therefore does not bear a sharp and unique  
relation to the bulk FQH state.
\end{abstract}

\maketitle

The interior of a fractional quantum Hall (FQH) system  [\onlinecite 
{Tsui}] is gapped, but
massless excitations exist at its edge, which constitutes a realization
of a one-dimensional electron liquid  described generically by the
Tomonaga-Luttinger theory [\onlinecite{GiulianiVignale}].  Much  
attention has been focused on the edge
physics since the work of Wen [\onlinecite{WenIntJModPhy}], where
it was conjectured that the exponent describing the long distance,  
low energy physics of the chiral (unidirectional)
FQH edge is a unique ``topological" quantum number for any given FQH  
state, independent of details, just as the quantum Hall resistance.   
Our understanding of the ordinary
one-dimensional liquids is largely based on the method of  
bosonization, which exploits
a one-to-one correspondence between the fermionic
and bosonic Fock spaces in one dimension, and identifies a relationship
between the operators on these spaces; specifically, the fermionic field
operator $\hat\psi(x)$ is related to the bosonic field operator $\hat 
\phi(x)$
through the expression $\hat\psi(x)\sim \exp[-i\hat\phi(x)]$, which  
can be
established  rigorously at the operator level
[\onlinecite{GiulianiVignale}].  In the absence of a similar rigorous  
derivation for the electron field operator at the edge of a
FQH system, Wen formulated an effective field theory approach (EFTA)  
[\onlinecite{WenIntJModPhy}] wherein he postulated that
the electron operator at the edge of the $1/m$ FQH state, defined by  
Hall resistance quantization at $R_H=h/(1/m)e^2$, is given by
\beq
\hat\psi(x)\sim e^{-i\sqrt{m}\hat\phi(x)}.
\label{Wenansatz}
\eeq
Antisymmetry under exchange quantizes $m$ to an odd integer value,  
independent of parameters other than the quantized Hall resistance,  
which leads to universal properties for the edge physics.
A direct test of this assertion is through tunneling of an
external electron laterally into the edge of FQH system.  For the
fractions $\nu=n/(2np+1)$, a generalization of Eq.~(\ref{Wenansatz})
predicts the $I$-$V$ characteristic
for electron tunneling into a Fermi liquid to be $I\sim V^\alpha$, where
  the tunneling exponent has a universal value of $\alpha=2p+1$.
Ingenious experiments
[\onlinecite{Grayson,Chang1,Chang2,Chang3}] have measured the edge  
exponent by determining the $I$-$V$
characteristics for tunneling from a three-dimensional Fermi liquid into
the FQH edge.  While they establish the existence of non-Fermi liquid  
(Tomonaga-Luttinger) behavior,  with an exponent different from a
one-dimensional Fermi liquid ($\alpha=1$), they also show discrepancy  
from the
EFTA prediction of Wen.  Specifically, experiments find an exponent  
that is continuously varying with the filling factor, and thus is not  
determined solely by the quantized Hall conductance.  Furthermore,  
the measured exponents are significantly different from the EFTA  
predictions: at filling factors $\nu=1/3$, $2/5$, and $3/7$, the  
exponents are $\sim$ 2.7,  $2.3$, and $2.1$, respectively [\onlinecite 
{Chang1,Grayson,Chang2,Chang3}], to be compared to the EFTA  
prediction of 3.0.

A number of theoretical papers have addressed this
inconsistency [\onlinecite{Chang3,Wan03,Yang03,Joglekar,ZulPalMacD,  
Orgad,Conti,  
conti98,Lopez,zuelicke99,levitov01,MandalJain,PalaciosMacDonald,  
Jolad}].
Some of these suggest
that the disagreement is due to edge reconstruction, which produces  
several
counter-propagating edge modes (for which the exponent is not universal)
[\onlinecite{ChamonWen,Wan03,Yang03,Joglekar,Orgad}], while some
  propose that the inconsistency
persists even in the absence of edge reconstruction, thus pointing to  
a more
fundamental deficiency of the EFTA [\onlinecite 
{MandalJain,ZulPalMacD}].  A resolution of this issue is important  
not only in its own right, but also in view of the potentially useful  
notion that the character of a bulk FQH state can be ascertained from  
the behavior of its edge physics [\onlinecite{Radu}], which logically  
rests on the existence of a unique relationship between the two and  
hence the universality of the latter.

\begin{figure}[htbp]
\begin{center}
\includegraphics[scale=0.22,viewport=0 62 550 408,clip]{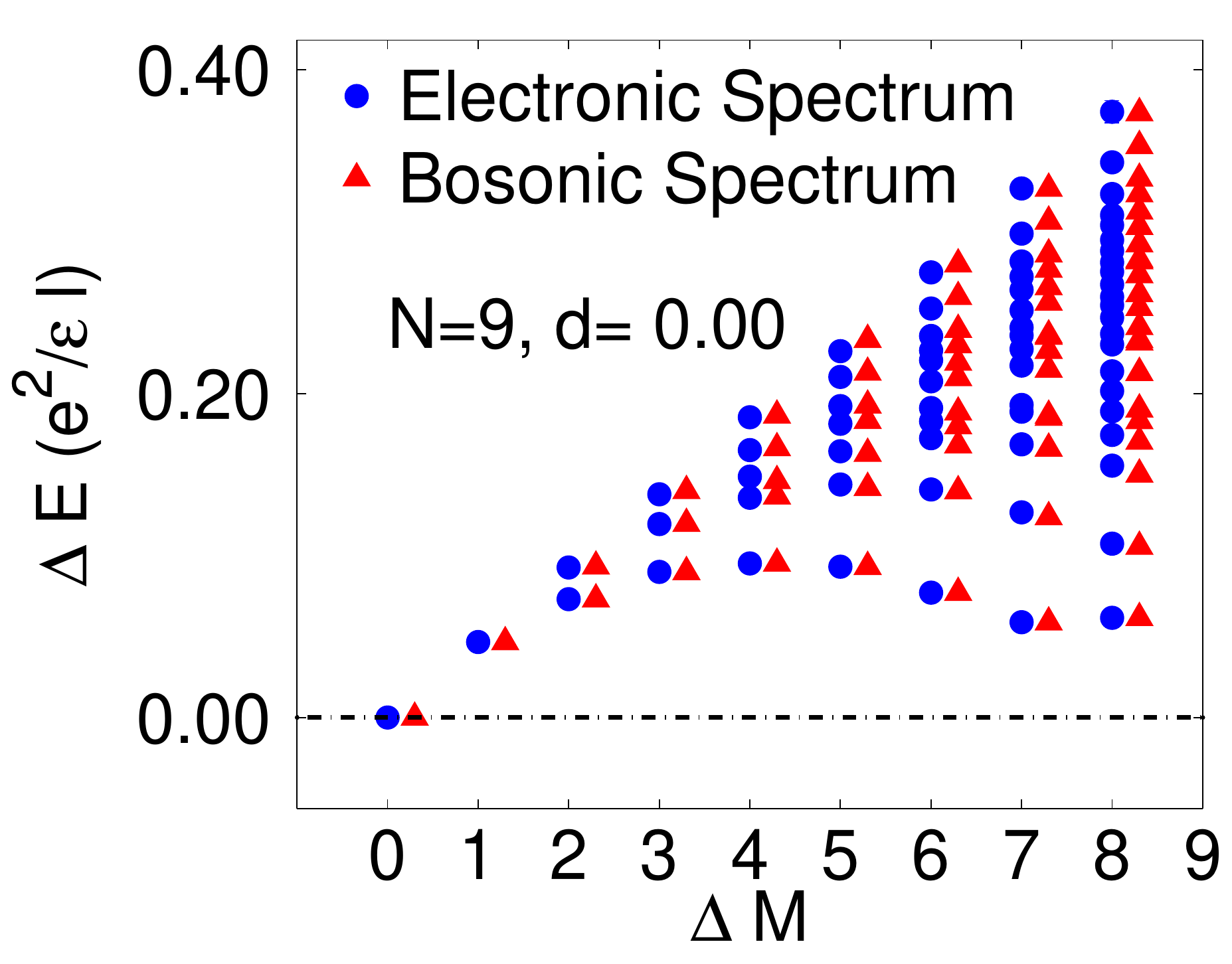}
\includegraphics[scale=0.22,viewport=0 62 550 408,clip]{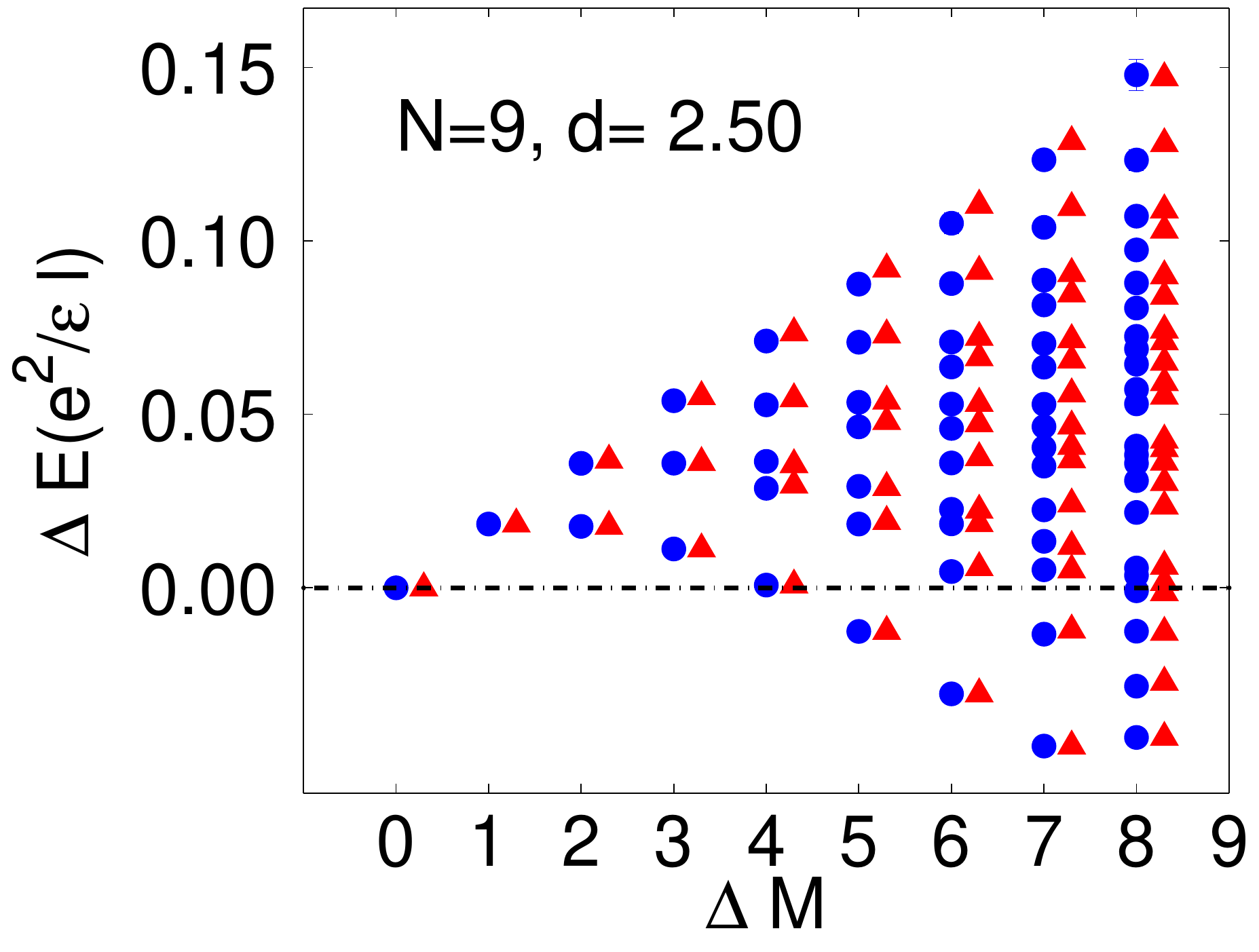}
\includegraphics[scale=0.22,viewport=0 62 550 408,clip]{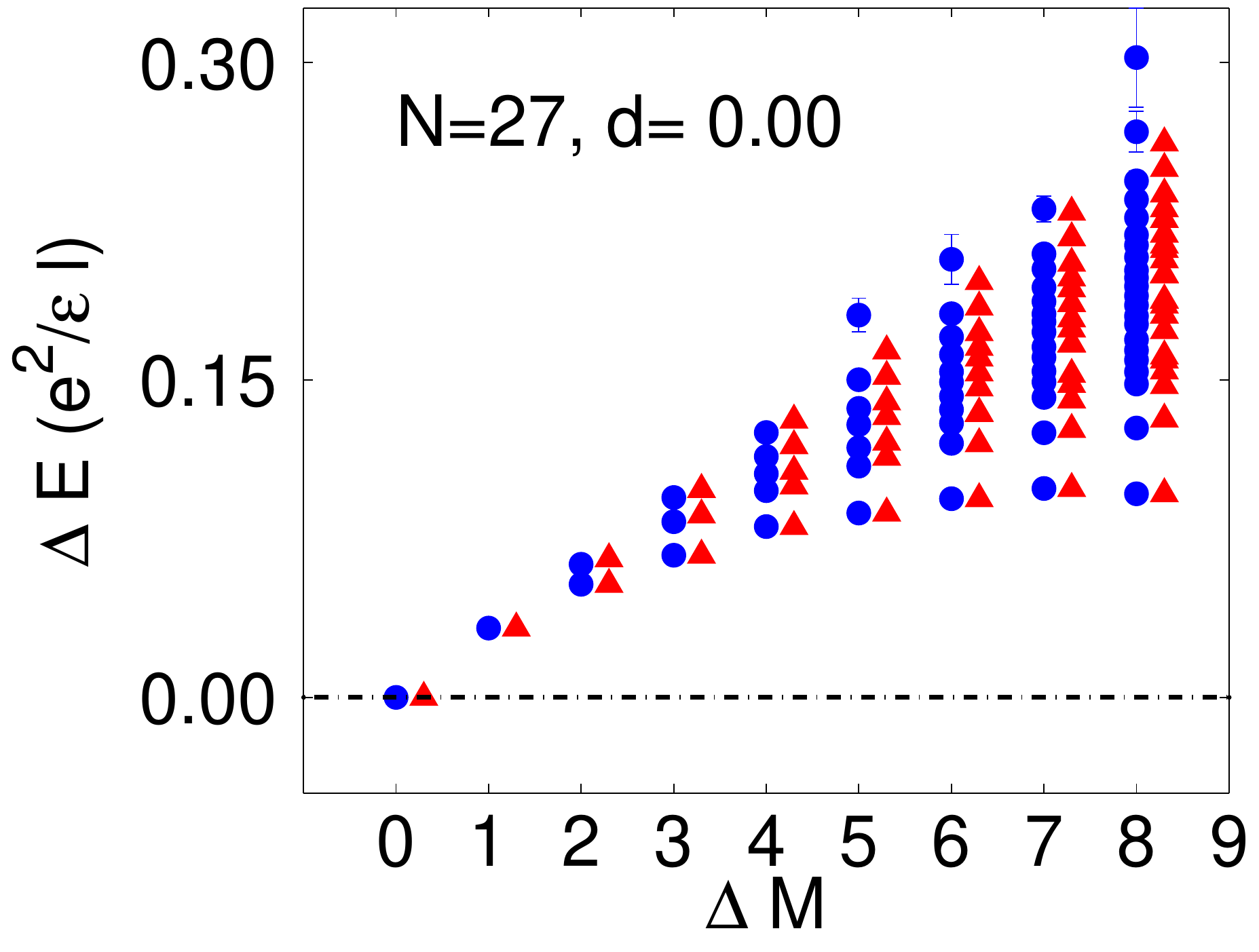}
\includegraphics[scale=0.22,viewport=0 62 550 408,clip]{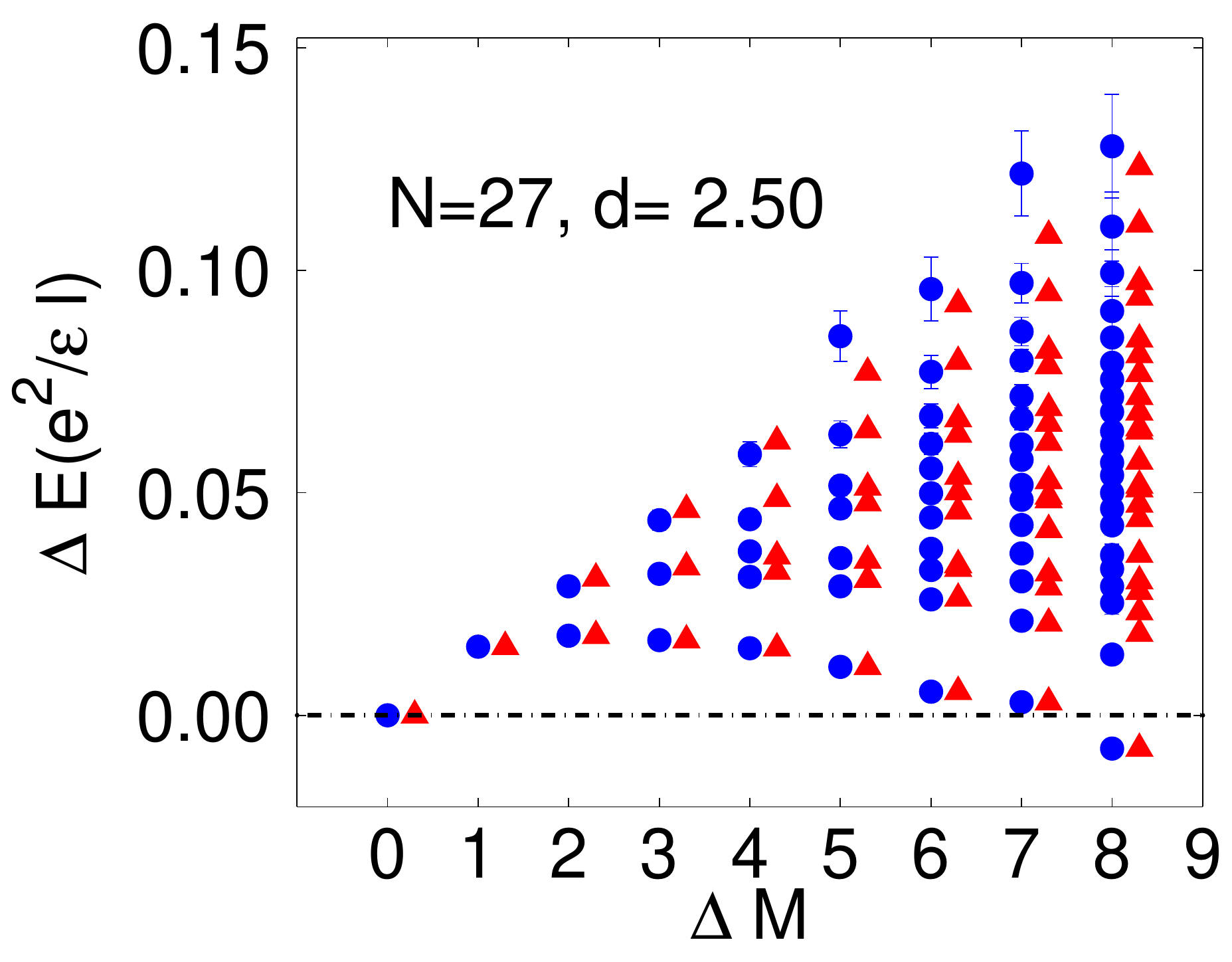}
\includegraphics[scale=0.22,viewport=0 0 550 408,clip]{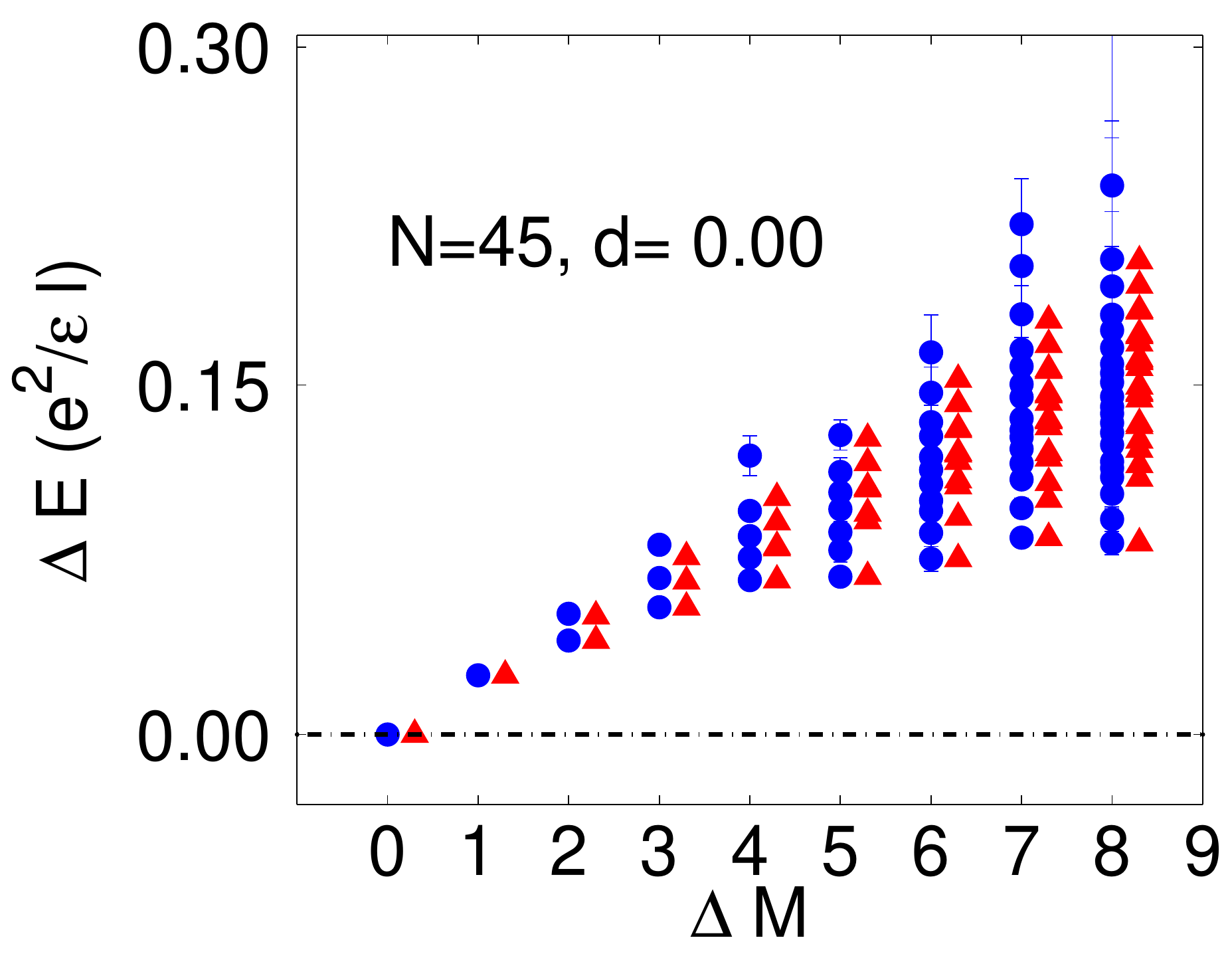}
\includegraphics[scale=0.22,viewport=0 0 550 408,clip]{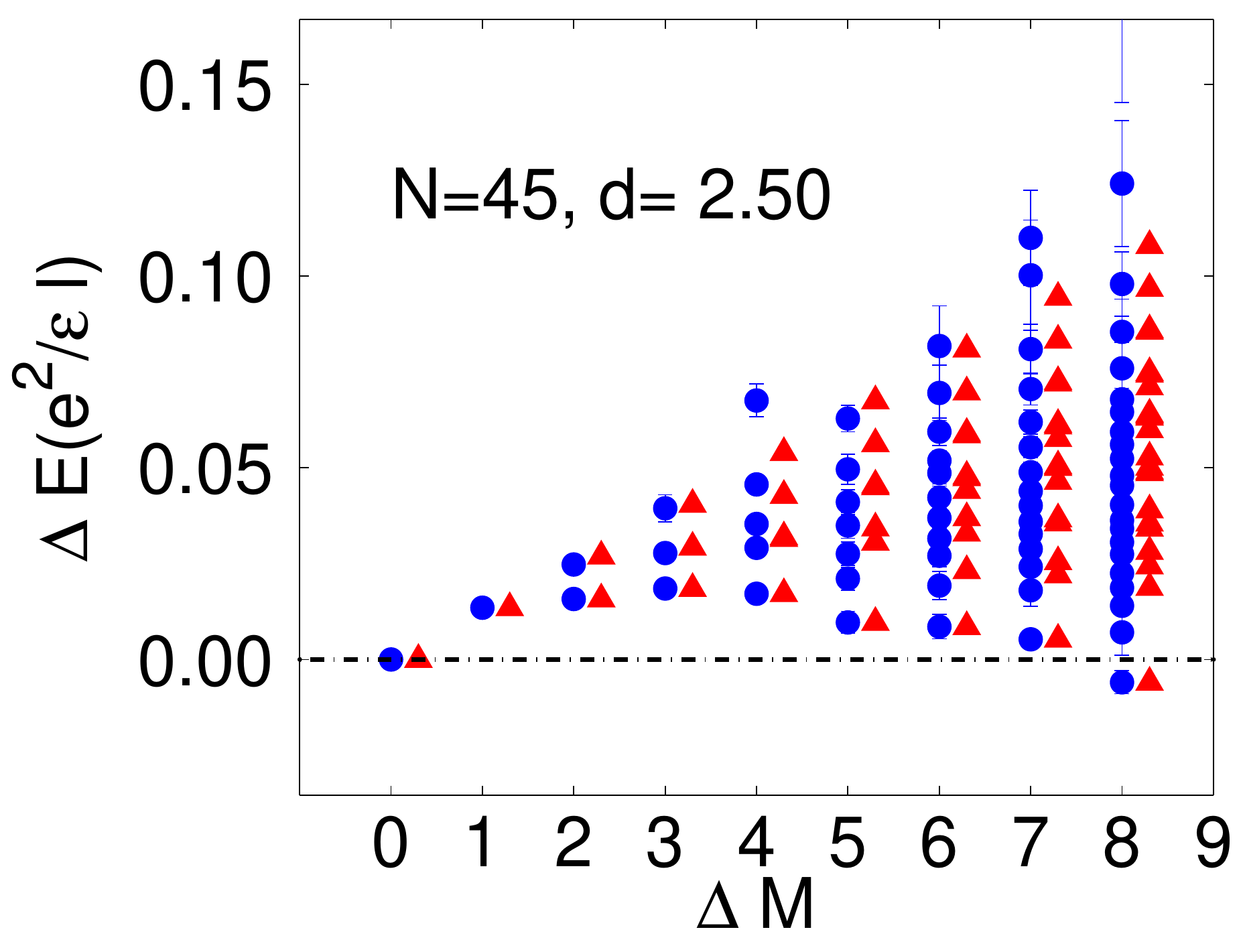}
\caption{[Color online]
Energy spectrum for the edge excitations of $\nu=1/3$ for $N=9,27,45$  
particles at electron-background separations in the range $d=0.0$ and  
$2.5$. Blue dots indicate the energies obtained by CF  
diagonalization, whereas the adjacent red triangles (shifted along the x axis for clarity)  show the bosonic  
spectra (see text for explanation). All energies are quoted in units of $e^2/\epsilon l$, and measured  
relative to the energy of the ground state at $\Delta M=0$.  $\Delta M 
$ is the angular momentum of the excited state.}
\label{FigEnTry}
\end{center}
\end{figure}

Exact diagonalization studies often provide an unprejudiced,  
reliable, and decisive tool for testing ideas in the field of the FQH  
effect.  For the edge physics, however, it has not been clear if the   
discrepancy between the finite system results and the EFTA is  
intrinsic or a finite size artifact; finite size corrections are more  
severe for the edge physics [\onlinecite{Jolad}] because of power law  
decay of correlations, in contrast to the Gaussian decay in the bulk.
Unfortunately,  the dimension of the Hilbert space grows  
exponentially with the number of electrons, making it impossible to  
increase the system sizes significantly in exact diagonalization  
studies.

In this Letter we report on substantial microscopic tests of the EFTA  
by diagonalizing the Coulomb Hamiltonian in a truncated space of edge  
excitations.  Specifically, we consider the edge excitations of the  
1/3 FQH state in the disk geometry, and the truncated space contains  
all states of the form
\beq
\Psi^M_\alpha =  \prod_{j<k}(z_j-z_k)^{2}
\Phi^{M^*}_\alpha, \;\; M=M^*+N(N-1),
\label{PsiAlpha}
\eeq
where $z_j=x_j-i y_j$ denotes the electron coordinates as a complex  
number, $M$ is the total angular momentum of the sate, and $\Phi^{M^*} 
_\alpha$ are all lowest Landau level states (labeled by $\alpha=1, 
\cdots, D^*$) at total angular momentum $M^*$.  The dimension of this  
basis space is much smaller than the full dimension of the lowest  
Landau level states at $M$, which makes it possible to investigate  
much larger systems; we have studied as many as 45 particles.
The restriction to this basis is equivalent to restricting composite  
fermions [\onlinecite{JainCFpaper}] to their lowest $\Lambda$ level  
(also known as composite fermion Landau level); the space of states  
can be enlarged in the standard manner [\onlinecite{CFD}] by also  
including at $M^*$ states occupying successively higher $\Lambda$  
levels (and projecting the total wave function onto the lowest Landau  
level), but that will not be necessary for our present purposes.  The  
lowest-$\Lambda$-level approximation for composite fermions is known  
to be excellent, and we have also confirmed its accuracy explicitly  
for edge excitations for systems with six and seven particles, for  
which exact results are available, both with and without the  
confinement potential.  Also, $\Psi^M_\alpha$ are the only states  
that survive if we add to the Coulomb interaction an appropriate  
infinitely strong short range interaction that annihilates states  
containing electronic pairs with angular momenta equal to unity; our  
results below are exact for this model.
Therefore, we believe that our truncated Hilbert space ought to  
capture the topological nature, if it exists, of the edge physics.   
We believe that this model actually gives the best chance for  
universal behavior; mixing with higher $\Lambda$ levels can only  
spoil it [\onlinecite{MandalJain}].

We consider a system of two dimensional electron gas in disk  
geometry. The neutralizing background has uniformly distributed  
positive charge contained in a disk $\Omega_N$ of radius $R_N=\sqrt 
{2N/\nu}$ for a system of $N$
particles at filling factor $\nu$;  the positively charged disk is  
separated by a distance $d$ from the electron disk (quoted in units  
of the magnetic length below). The electrons are approximately  
confined to the same radius because of charge neutrality in the  
interior.
This system is modeled by the following realistic Hamiltonian:
\beq
H=E_{\rm K}+V_{\rm ee}+V_{\rm eb}+V_{\rm bb}
\eeq
where the terms on the right hand side represent the kinetic,  
electron-electron, electron-background,
and background-background Coulomb interaction
energies, respectively.
At large magnetic
fields only the lowest Landau level states are occupied, hence the  
kinetic
energy $\hbar \omega_c/2$ (where $\omega_c \equiv eB/m_b c$ is the  
cylcotron frequency) is a constant and will not be considered  
explicitly.

The wave functions $\Psi_{\alpha}^M$ are in general not orthogonal,  
and we use the method of composite-fermion diagonalization (CFD)  
[\onlinecite{CFD}] to orthogonalize them by the Gram-Schmidt  
procedure, evaluate the Hamiltonian matrix elements, and diagonalize  
it to obtain the eigenvalues and eigenvectors.  All matrix elements  
and scaler products needed for this purpose are evaluated by the  
Monte Carlo method, as explained elsewhere in the literature  
[\onlinecite{CFD}].  While sufficiently accurate energy spectrum  
requires $\sim$10-20 million Monte Carlo iterations, the spectral  
weights require $\sim$ 200 million iterations for each eigenstate.   
These numbers do not vary significantly with $N$, but the computation  
time increases exponentially with $N$ and $\Delta M$, limiting our  
study to systems with $N=45$ for energy, and to $N=27$ for spectral  
weights.  The energies were calculated for $\Delta M=1-8$ and the  
spectral weights for $\Delta M = 1-4$.

Using the CF diagonalization procedure, we compute the spectra of  
edge excitations of the $1/3$ state, shown in Fig.~(\ref{FigEnTry}),  
for several parameters in the range $N=6$-$45$, $d=0$-$2.5$, and $ 
\Delta M =0$-$8$.  Our large system calculations confirm an earlier  
study [\onlinecite{Wan03}] that
edge reconstruction occurs for $d$ larger than a critical separation,  
which is approximately 1.5-2.0 magnetic lengths.
For $d$ greater than the critical separation, a simple explanation  
for the observed nonuniversality follows in terms of edge  
reconstruction; a model that assumes a single chiral mode is  
inadequate to describe experiments, forcing one to consider multiple  
edge modes, which produces nonuniversal results.
However, the important question remains whether universality occurs  
in the absence edge reconstruction.

We address this issue by following the pioneering work of Palacios  
and MacDonald[\onlinecite{PalaciosMacDonald}] to test the validity of  
Eq. (\ref{Wenansatz}), upon which the notion of universality rests.
Specifically, we compare certain matrix elements of the electron  
field operator, computed from our CF diagonalization results, with  
the predictions of the bosonized form in Eq. (1).  We also consider $d 
$ greater than the critical separation for completeness; here, we
assume that the ground state remains at $\Delta M=0$, which can be  
arranged by adding an ad hoc angular momentum dependent single  
particle energy term that strongly penalizes the edge excitations  
responsible for edge reconstruction, but does not change either the  
eigenfunctions or the energy ordering of states at a given $\Delta M 
$.  (This can be accomplished by adding an appropriate parabolic  
confinement term which adds to the total energy a term proportional  
to the total angular momentum.)

The spectral weights are defined by
\beq
	C_{\{n_l\}}= \frac{\langle \{n_l\} | \hat\psi^\dagger(\theta)|0\rangle}
	{\langle 0 | \hat\psi^\dagger(\theta)|0\rangle},
	\label{spectralweight}
\eeq
where $|\{n_l\}\ra$ represents the bosonic state with occupation $\{n_l\}$, $|0\ra$ is the vacuum state with zero bosons, and $\hat\psi^\dagger(\theta) $ is the electron creation operator at position $\theta$ along the edge circle. Here $l$ denotes single boson angular  
momentum; the total angular momentum is denoted by
$\Delta M= \sum_l  l \,n_l$ and the total energy by $\Delta E=\sum_l  
n_l \epsilon_l$, with $\epsilon_l$ being the energy of a single boson  
at angular momentum $l$.
With the help of
$\hat{\psi}^\dagger (\theta) \propto  e^{-i\sqrt{m}\hat\phi(x)}=\sqrt 
{z} e^{-i\sqrt{m}\hat{\phi}_+(\theta)}
e^{-i\sqrt{m}\hat{\phi}_-(\theta)}$,
$\hat{\phi}_+(\theta)=-\sum_{l>0} (1/\sqrt{l}) a_l^\dagger e^{il 
\theta}=\hat{\phi}_-^{\dagger}(\theta)$,
  it is straightforward to obtain the predictions for the spectral  
weights:
\beq
|C_{\{n_l\}}|^2=\frac{m^{n_1+n_2+\cdots}}{n_1!n_2!\cdots 1^{n_1} 2^ 
{n_2}\cdots}
\eeq
We note that the denominator in Eq.
(\ref{spectralweight}) eliminates the unknown normalization constant
$\sqrt{z}$.

\begin{figure}[htbp]
\begin{center}
\includegraphics[scale=0.22,viewport=0 58 550 410,clip]{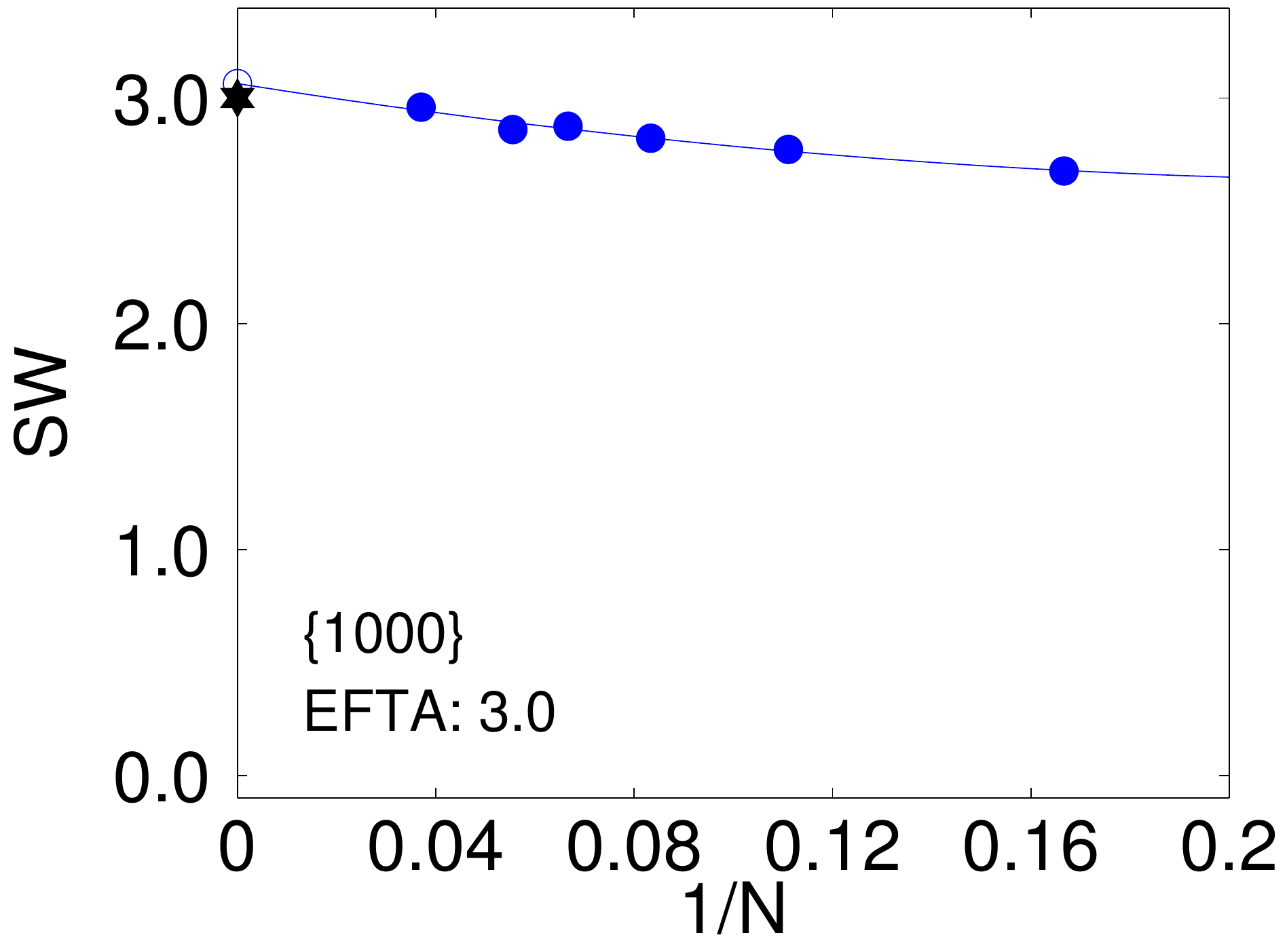}
\includegraphics[scale=0.22,viewport=0 58 550 410,clip]{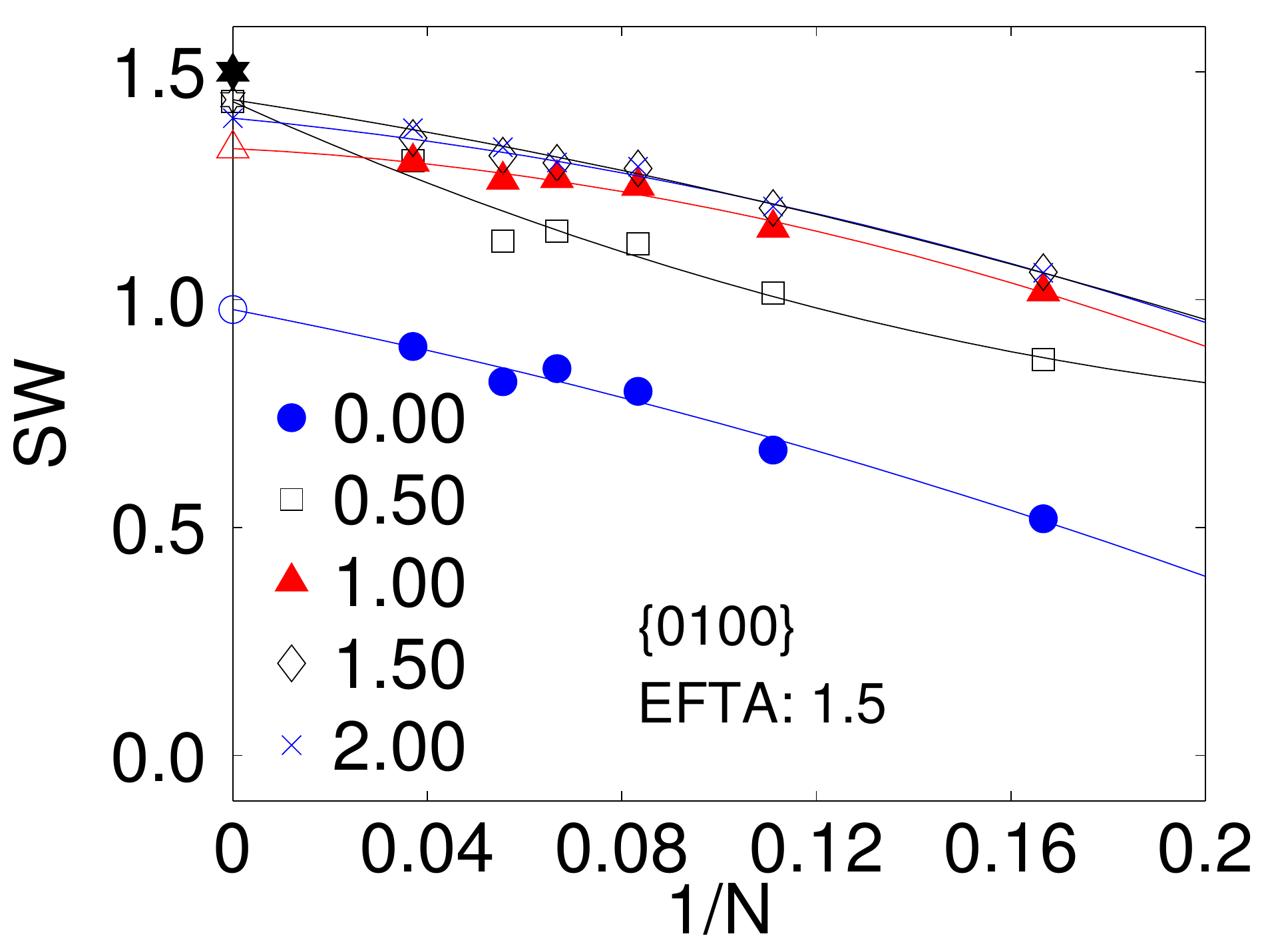}
\includegraphics[scale=0.22,viewport=0 58 550 410,clip]{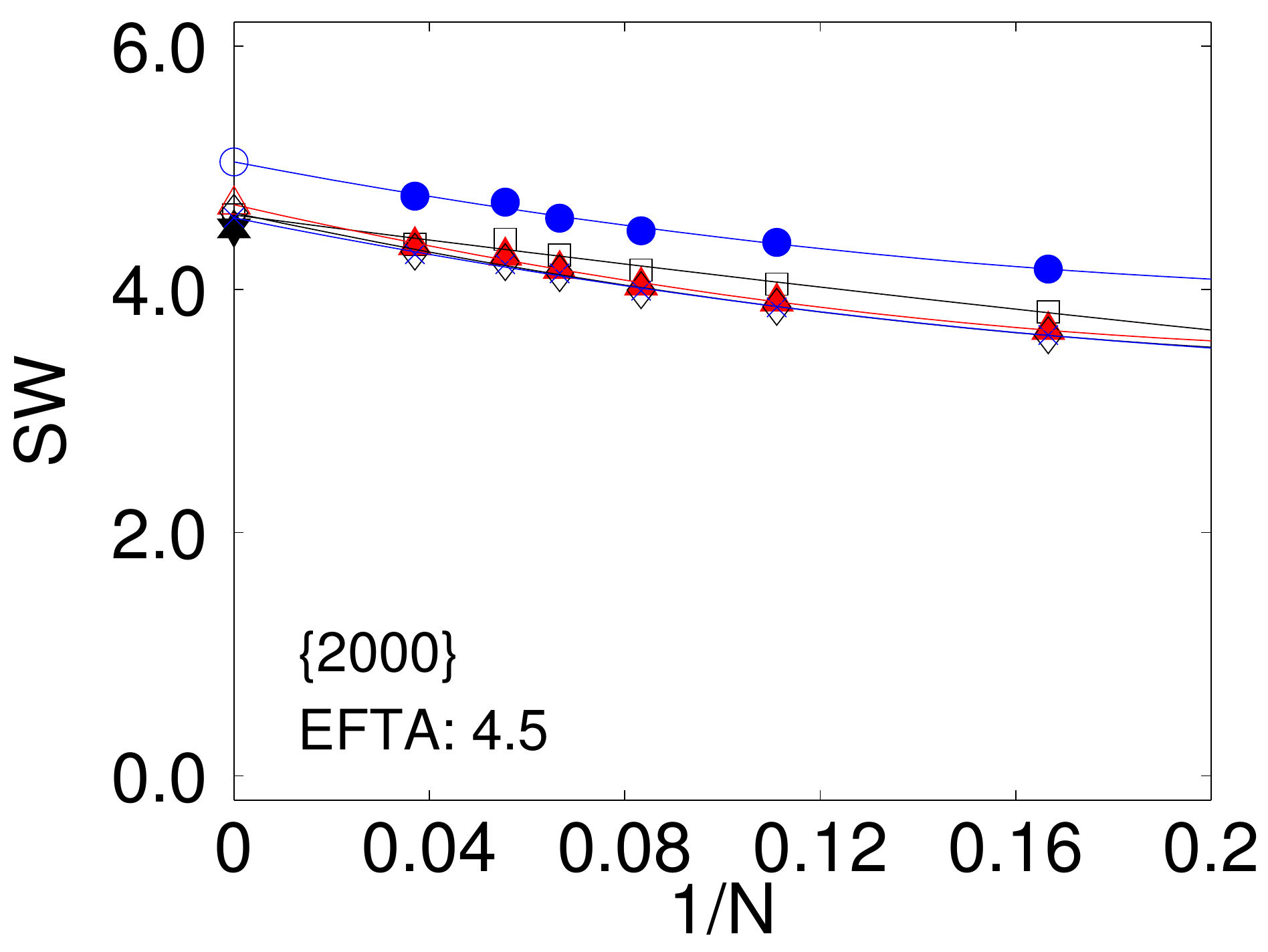}
\includegraphics[scale=0.22,viewport=0 58 550 410,clip]{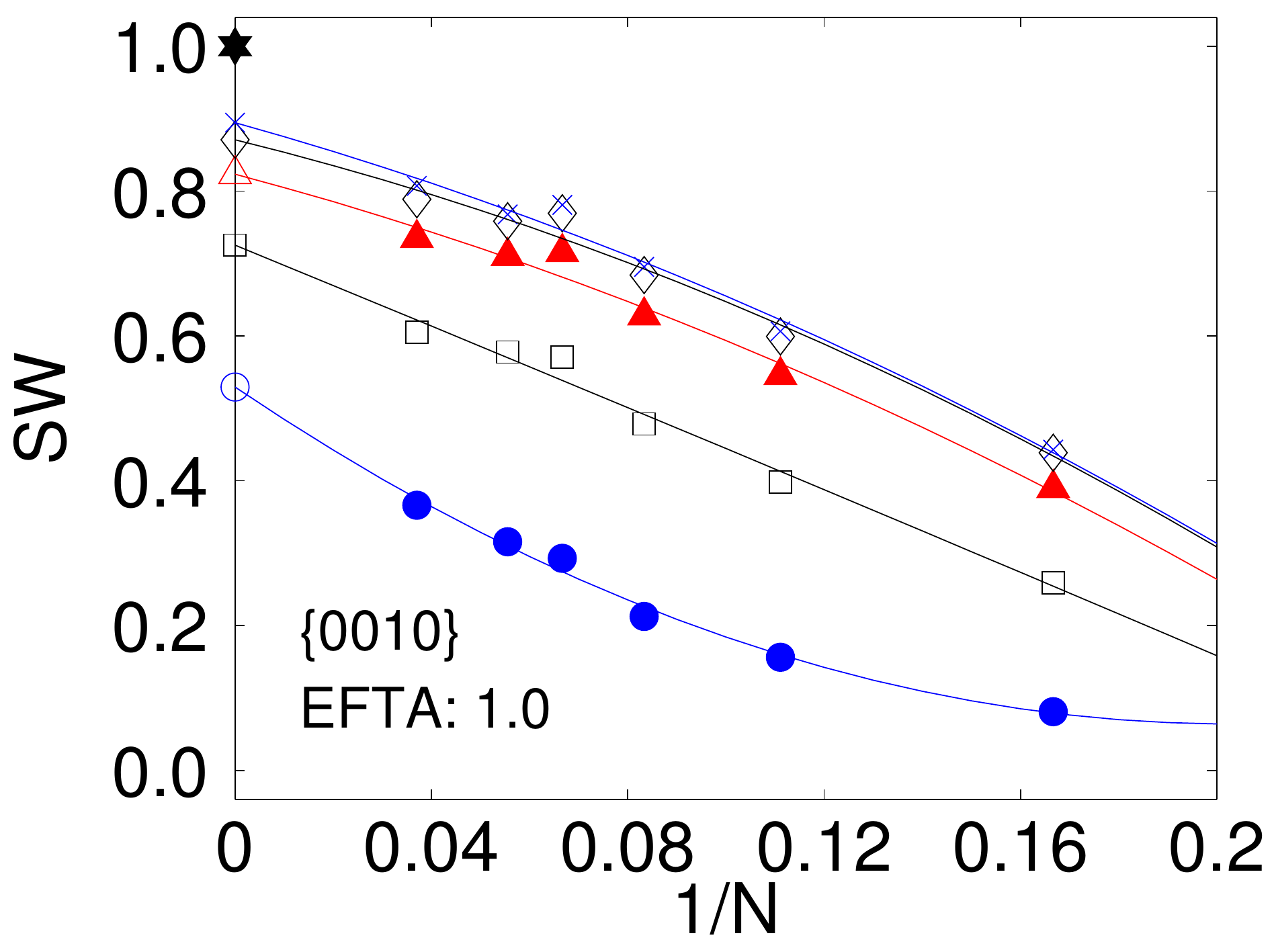}
\includegraphics[scale=0.22,viewport=0 58 550 410,clip]{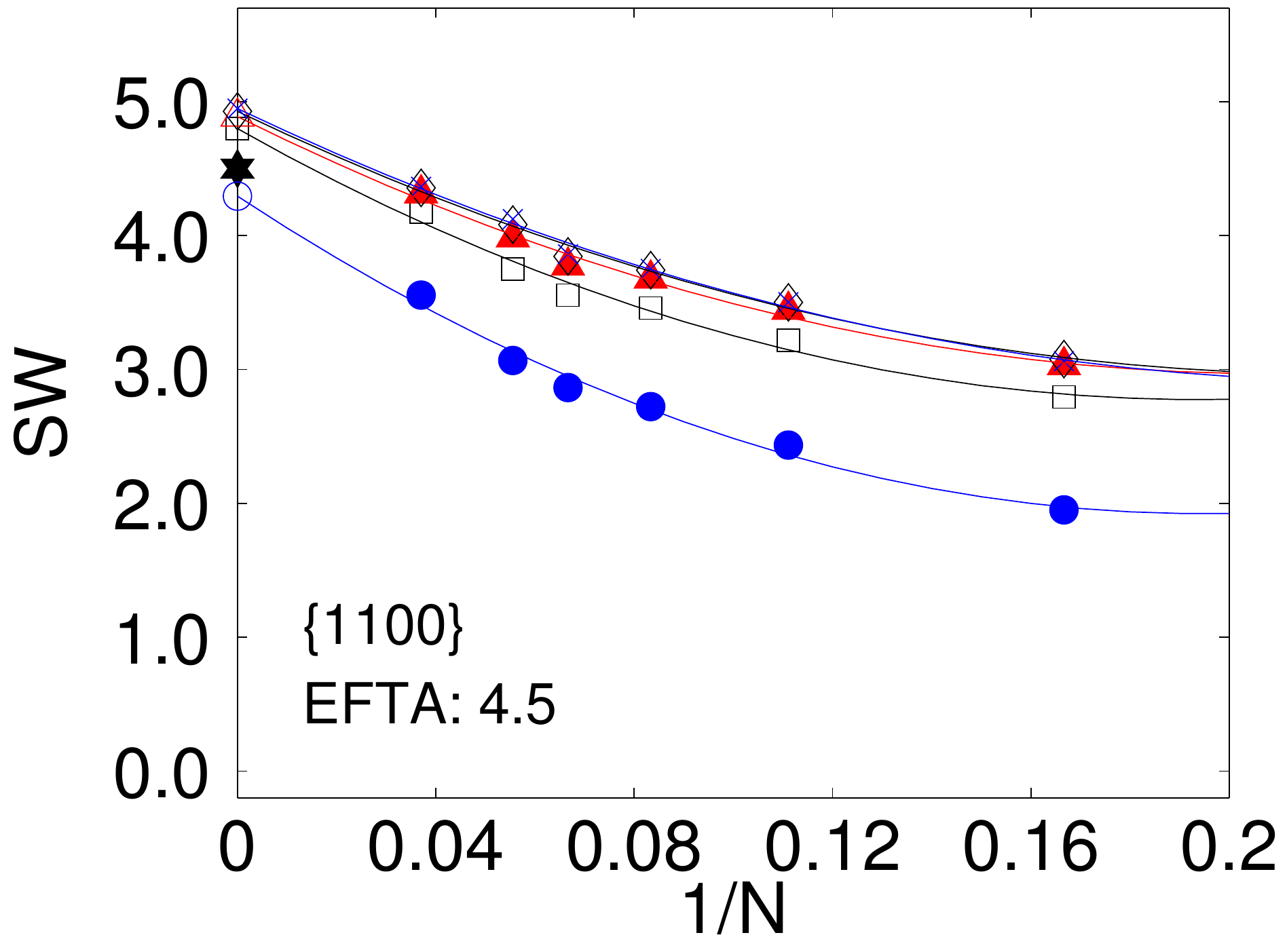}
\includegraphics[scale=0.22,viewport=0 58 550 410,clip]{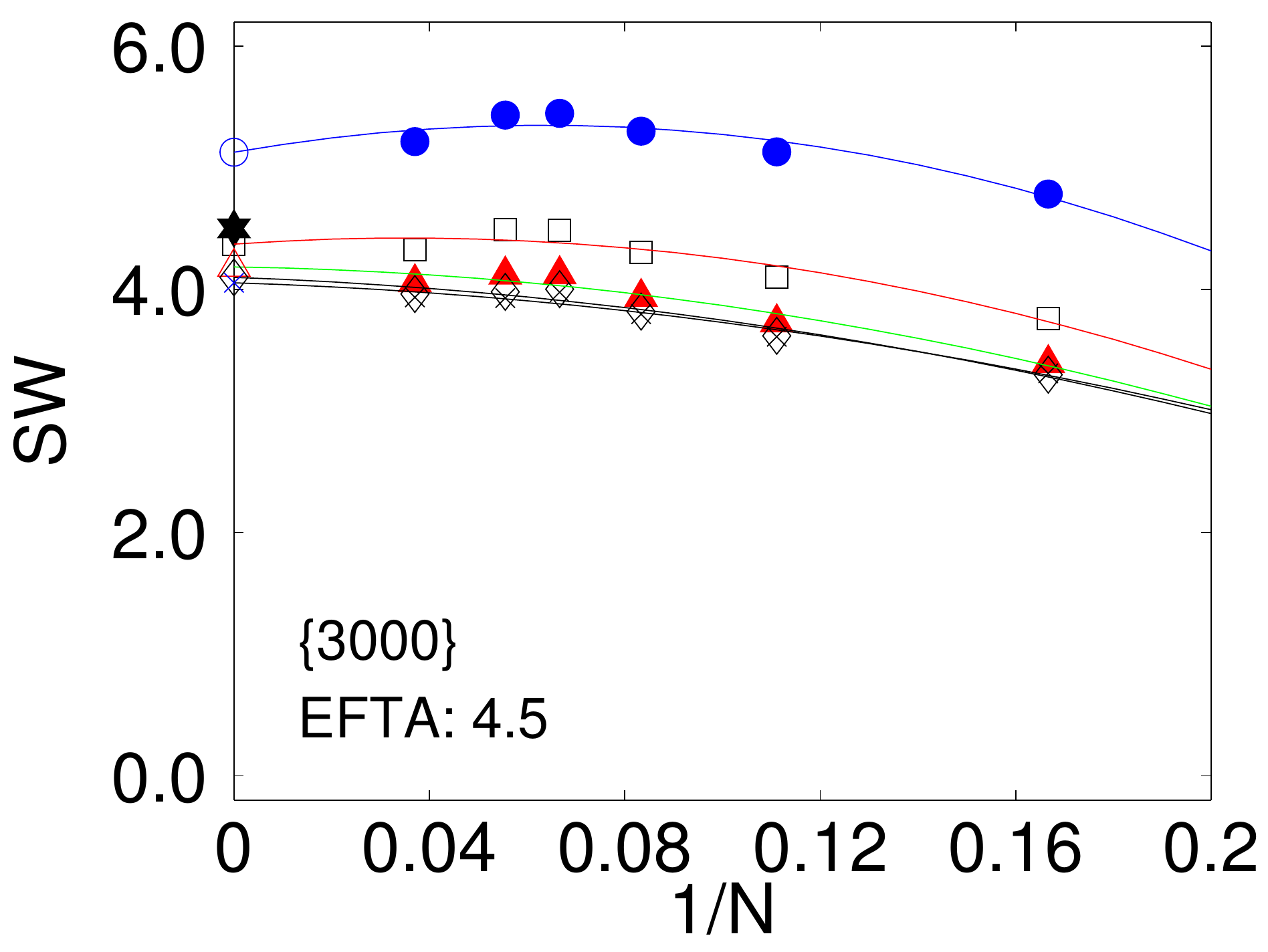}
\includegraphics[scale=0.22,viewport=0 58 550 410,clip]{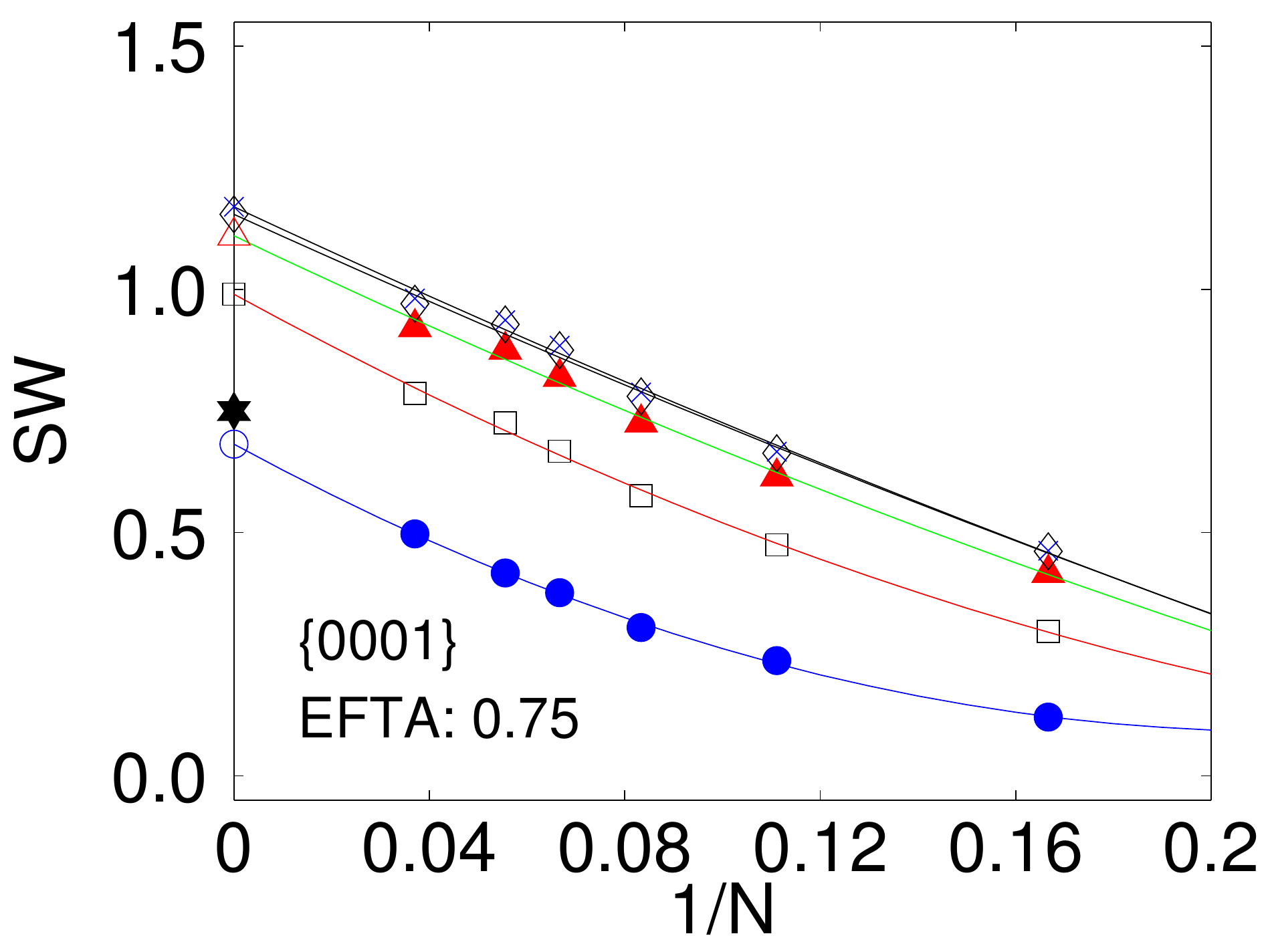}
\includegraphics[scale=0.22,viewport=0 58 550 410,clip]{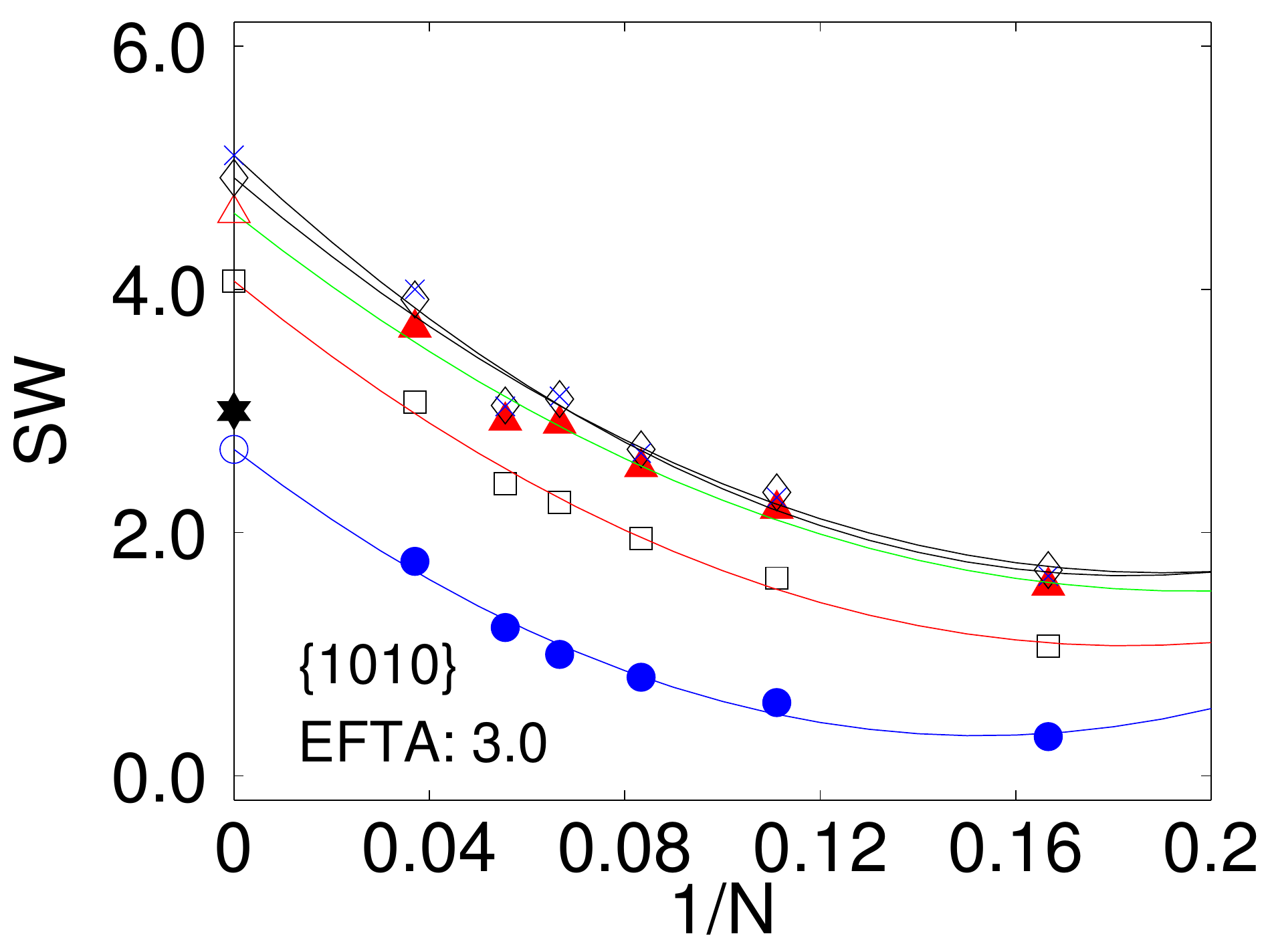}
\includegraphics[scale=0.22,viewport=0 0  550 410,clip]{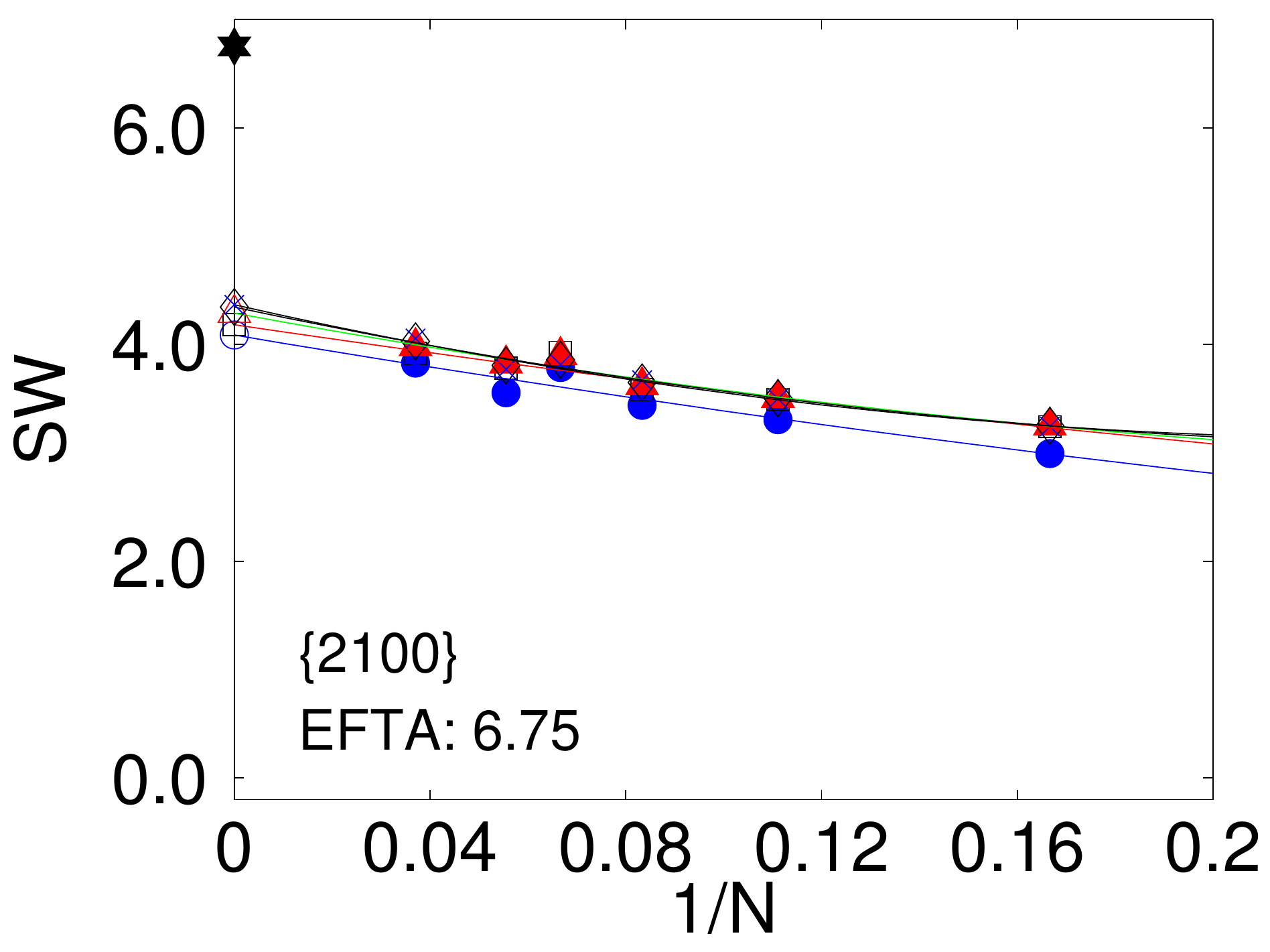}
\includegraphics[scale=0.22,viewport=0 0  550 410,clip]{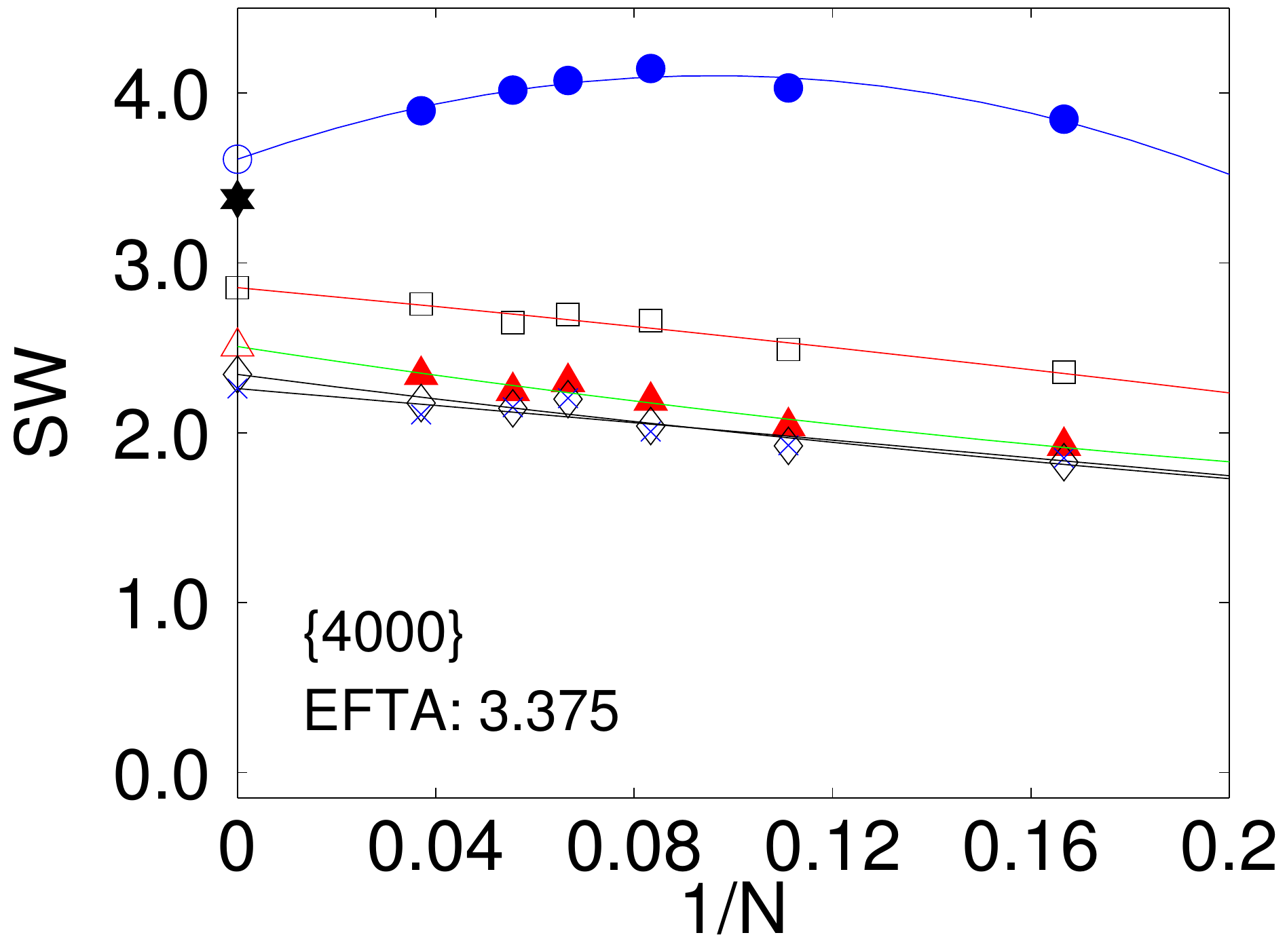}
\caption{[Color online] $N$ dependence of spectral weights $|C_{\{n_l 
\}}|^2$ for several states ${\{n_l\}}$, indicated on each panel, and  
various separations $d$ (quoted in units of the magnetic length in  
the $\{0100\}$ panel). The EFTA prediction from Eq.~(1)  is indicated  
by a star on the y-axis, with the value also given on each panel.   
The points on the y-axis are determined by a quadratic fit to the  
finite $N$ results.}
\label{FigSW1}
\end{center}
\end{figure}

In order to obtain the spectral weights from our electronic spectra,  
it is natural to identify the vacuum state $|0\rangle$ with the  
ground state of
interacting electrons at $\nu=1/m$, denoted by $|\Psi_0^N \rangle$,
and the field operator has the standard
meaning of
$
\hat\psi^\dagger(\theta)=\sum_l\eta_l^*(\theta) c_l^\dagger \equiv  
\sum_l
\psi_l^\dagger(\theta)
$,
where $c_l^\dagger$ and $c_l$ are creation and annihilation operators  
for an
electron in the angular momentum $l$ state.
The denominator of Eq. (\ref{spectralweight}) corresponds to
\beq
\langle 0|\hat\psi^\dagger(\theta)|0\rangle = \frac{\langle \Psi_0^{N+1}
|\hat\psi^\dagger_{L_0}(\theta)|\Psi_0^N\rangle}{\sqrt{\langle \Psi_0^ 
{N+1}|\Psi_0^{N+1}\rangle \langle\Psi_0^N |\hat\psi_{L_0}(\theta) \hat 
\psi^\dagger_{L_0}(\theta)|\Psi_0^N\rangle  }} ,
\eeq
where $|\Psi_0^N \rangle$ is the ground state of $N$ interacting  
electrons
at $\nu=1/m$, and $L_0=mN$.  The numerator is similarly defined as as
\beq
\langle {\{n_l\}} |\hat\psi^\dagger(\theta)|0\rangle = \frac{\langle
\Psi_{\{n_l\}}^{N+1}|\hat\psi_L^\dagger(\theta)|\Psi_0^N\rangle}{\sqrt 
{\langle
\Psi_{\{n_l\}}^{N+1}|\Psi_{\{n_l\}}^{N+1}\rangle \langle\Psi_0^N |\hat 
\psi_{L}(\theta) \hat\psi^\dagger_{L}(\theta)|\Psi_0^N\rangle  }}.
\label{eq9}
\eeq
Here, we have
\beq
\hat\psi_L^\dagger |\Psi_{0}^N\rangle=\mathcal{N}_L\mathcal{A}\left[
z_{N+1}^Le^{-|z_{N+1}|^2/4}\Psi_0^N(z_1,z_2\dots,z_N)\right],
\label{ElectronCreate}
  \eeq
where ${\cal A}$ is the antisymmetrization operator, ${\cal{N}}_L$ is  
the
normalization constant, and $L\equiv L_0+\Delta M$ is the angular  
momentum of added  electron.

The wave function $\Psi_{\{n_l\}}^{N+1}$, the electronic counterpart  
of the
bosonic state $|\{n_l\}\rangle$, clearly represents an excited state  
at total angular momentum $M=\Delta M+mN(N+1)/2$, which should also  
be related to the total angular momentum of the $N$ particle ground  
state through $M=L+mN(N-1)/2$.
At each $\Delta M$, there are in general many eigenstates.  Following  
Ref.~[\onlinecite{Wan03}] we identify $\epsilon_l$, the energy of a  
single boson with angular momentum
$l$, with the lowest energy at $l=\Delta M$ in the calculated spectrum.
Using the equations $\sum_l l n_l =\Delta M$ and $E_{\{n_l\}}=\sum_l n_l
\epsilon_l$, the energies of the all bosonic states $\{n_l\}$ can now  
be obtained
(see Fig. \ref{FigEnTry}), which can then be identified with the  
corresponding electronic states. We note that in the lowest $\Lambda$  
level subspace, the numbers of electronic and bosonic states are  
equal at each $\Delta M$, so a one to one correspondence between the  
two sets of states can be established from their energy orderin.
For small systems (for example, $N=9$ in Fig.~\ref{FigEnTry}), the  
CFD spectra  and the bosonic spectra are very close to each other,  
which explicitly confirms the interpretation of the lowest branch as  
the single boson branch.  The agreement between the electronic and  
bosonic spectra becomes less accurate with increasing $N$ or $\Delta M 
$, but still remains adequate for the low energy states, which will  
be our focus. (The higher energy states of the spectra shown in  
Fig.~1 mix with higher $\Lambda$ level excitations of composite  
fermions, not considered in our model.)

Figure~(\ref{FigSW1}) shows the squared spectral weights for  
different excited states as a function of $N$ and $d$. A quadratic  
fit extrapolates the result to the thermodynamic limit $1/N=0$.  The  
EFTA predictions from Eq.~(1) are also shown in each panel.  These  
plots demonstrate the central result of our work: the spectral  
weights are nonuniversal; they depend on $d$; and they do not  
extrapolate to the EFTA value. For the $\{ 1000 \}$ excitation the  
thermodynamic result agrees with predicted result of $3.0$ for all $d 
$; however, in this case our truncated space contains a single  
(center-of-mass) excitation, the wave function for which is  
independent of interactions (within our model), and therefore the  
agreement is not meaningful.  For many cases, the deviation from the  
EFTA value is substantial; even the spectral weights of single boson  
states, such as $\{0100\}$ and $\{0010\}$ exhibit significant $d$  
dependence.

Our study thus indicates that Eq.~1 is not valid for the $1/m$ FQH  
edge for the Coulomb interaction, and therefore there is no reason to  
expect the edge exponent to be a  topological quantum number for the  
$1/m$ FQH state; this conclusion very likely holds for other FQH  
states as well, given that their edges are believed to be more  
complex.  The problem of how in reality the electron field is related  
to the bosonic field remains unresolved, however.  Following Ref.~ 
[\onlinecite{ZulPalMacD}] one may abandon the antisymmetry  
requirement and try an expression of the type $\hat\psi(x)\sim e^{-i 
\sqrt{\alpha}\hat\phi(x)}$ with arbitrary $\alpha$; we have found  
that no single value of $\alpha$ gives a satisfactory description of  
all spectral weights that we obtain from numerical diagonalization.   
It is also worthwhile here to mention the possibility of a nonlocal  
relation between the two, as suggested in Ref.~[\onlinecite 
{MandalJain}].

Other models for edge confinement have been used.  One such model  
[\onlinecite{Wan03}] restricts the single particle angular momentum  
to a maximum value of $l_{\rm max}=3(N-1)+l_0$, which may be a  
reasonable approximation for cleaved edge overgrowth [\onlinecite 
{Chang3}].  However, angular momentum conservation shows (and our  
Monte Carlo calculations explicitly confirm) that in this model the  
actual spectral functions identically vanish for $\Delta M > l_0$,  
resulting in an even more substantial disagreement with the EFTA  
predictions.

In summary, our study allows an estimate of certain thermodynamic  
properties of the FQH edge for a realistic Coulomb model, and makes  
what we believe to be a compelling case that the behavior at the FQH  
edge is intrinsically nontopological, as also demonstrated by  
experiments.  This result has obvious implications for the program of  
determining the nature of a bulk FQH state by probing its edge.

We acknowledge Paul Lammert, Diptiman Sen, and Chuntai Shi for  
insightful discussions and support
with numerical codes. The computational work was done on the LION-XO  
and LION-XC cluster of High Performance Computing (HPC) group,  
Pennsylvania State University.

\end{document}